\documentstyle[12pt,psfig]{article}
\newcommand{\beq}{\begin{equation}}
\newcommand{\eeq}{\end{equation}}
\newcommand{\beqa}{\begin{eqnarray}}
\newcommand{\eeqa}{\end{eqnarray}}
\def\nue{{\nu_e}}
\def\anue{{\bar\nu_e}}
\def\numu{{\nu_{\mu}}}
\def\anumu{{\bar\nu_{\mu}}}

\def\anutau{{\bar\nu_{\tau}}}
\newcommand{\dm}{\mbox{$\Delta{m}^{2}$~}}
\newcommand{\st}{\mbox{$\sin^{2}\theta$~}}
\begin{document}
\begin{center}
{\large{ \bf { MSW mediated neutrino decay and the solar 
neutrino problem}}}
\vskip 10pt
{\it Abhijit Bandyopadhyay\footnote{abanerjee@tnp.saha.ernet.in},  
Sandhya Choubey\footnote{sandhya@tnp.saha.ernet.in}, 
Srubabati Goswami\footnote{sruba@tnp.saha.ernet.in}}\\
\vskip 6pt
Saha Institute of Nuclear Physics,\\1/AF, Bidhannagar,
Calcutta 700 064, INDIA.\\
\end{center}
\vskip 30pt

\begin{center}
{\bf Abstract}
\end{center}

We investigate the solar neutrino problem assuming simultaneous presence of 
MSW  transitions in the sun and neutrino decay on the way from sun to 
earth.
We do a global $\chi^2$-analysis of the data on total rates in Cl, 
Ga and 
Superkamiokande (SK) experiments and the SK day-night spectrum data
and determine
the changes in the allowed region in the $\dm - \tan^2\theta$
plane in presence of decay.  
We also discuss the implications for unstable neutrinos in the  
SNO experiment.  

\vskip 15pt

\newpage
\section{Introduction}
The global analysis of the total rates measured in the Cl, Ga \cite{solar}
and 
SK experiments and the day-night spectrum data of SK \cite{sk1117} 
indicate that the large mixing angle MSW solution gives the
best description of the solar neutrino data \cite{carlos}.
However before a particular solution can be established one should 
rule out other possibilities.
In this spirit people have considered various non-standard neutrino 
properties and their implication for the solar neutrino problem \cite{other}. 
In this paper we consider a scenario where neutrinos are allowed to 
decay on their way from sun to earth
after undergoing MSW transitions in the sun. Such a possibility was 
discussed earlier in \cite{prag}. 

We consider two flavor mixing between $\nu_e$ and 
$\nu_\mu$/$\nu_\tau$ with the mass eigenstates $\nu_1$ and $\nu_2$. 
We assume that the 
heavier mass state $\nu_2$ is unstable, while 
the lighter neutrino mass state $\nu_1$ has lifetime much greater than the 
sun-earth transit time and hence can be taken as stable. 
There are two possible non-radiative decay modes\footnote{Radiative 
decays are severely constrained\cite{fukugita}.} 
\begin{itemize}
\item Model 1: If neutrinos are Dirac particles 
one has the decay channel $\nu_2 \rightarrow \bar{\nu}_{1R} + \phi$, 
where ${\bar{\nu}}_{1R}$ is a right handed singlet and $\phi$ is an
iso-singlet scaler. 
Thus all the final state particles 
for this model are sterile and there is no distinct signature 
of this decay apart from in disappearance experiments. 
This model is discussed in \cite{app}. In this model a light scalar boson 
$\phi$ with lepton number -2 and a singlet right handed 
neutrino is added to the 
standard model. The neutrino coupling to this scalar boson is given by
$g_{21} \nu_{R_1}^T C^{-1} \nu_{R_2}$, 
$C$ being the charge conjugation operator. 

\item Model 2: If neutrinos are Majorana particles, 
the decay mode is $\nu_2 \rightarrow \bar{\nu}_1 + J$, 
where J is a Majoron, produced as a result of spontaneous
breaking of a global $U(1)_{L_e - L_\mu}$ symmetry \cite{ajp}. 
In this model the neutrino masses are generated by extending 
the higgs sector of the standard model. 
Though the original triplet majoron model proposed by 
Gelmini and Roncadeli \cite{gm} is ruled out from the LEP 
data on Z decay to invisible
modes \cite{lep}, the model discussed in \cite{ajp}, which 
needs two additional triplet and one singlet scalar boson 
in the theory, can avoid the conflict with the LEP data, 
and at the same 
time predict
fast enough neutrino decay necessary to solve the solar 
neutrino problem.
In this model the $\bar{\nu}_1$ can be 
observed as a $\bar{\nu}_e$ with a probability $|U_{e1}|^2$ and as 
a $\anumu/\anutau$ with a probability $|U_{e2}|^2$. 
\end{itemize}

\noindent
In both the decay 
scenarios the rest frame lifetime of $\nu_{2}$ is given by \cite{acker}
\begin{equation}
\tau_{0} = \frac{16 \pi}{g^2} \frac{m_{2} (1 + m_{1}/m_{2})^{-2}}{\dm}
\label{tau0}
\end{equation}
where $g$ is the coupling constant, $m_i$ is the $\nu_i$ mass and 
$\Delta m^2 = m_2^2 - m_1^2$. 
Assuming
$m_{2} >> m_{1}$  the equation (\ref{tau0}) can be written as
\begin{equation}
g^2 \dm \sim 16 \pi \alpha
\label{galpha}
\end{equation}
where $\alpha$ is the decay constant related to $\tau_0$ as 
$\alpha = m_{2}/\tau_{0}$.
If we now incorporate the bound $g^2 < 4.5 \times 10^{-5}$ as obtained
from K decay modes \cite{barger}
we get the bound $\Delta m^2 > 10^{6} \alpha$ 
\footnote{Since the $m_{ee}$ element of the mass matrix is zero in the 
majoron decay model considered, the 
$0\nu$-majoron $\beta \beta$ decay does not take place in this 
model \cite{ajp} and the more stringent bound on g from  majoron emission 
in $\beta \beta$ decay \cite{bb} is not applicable.}. Since for  
a typical neutrino energy of 
10 MeV, one starts getting decay effects over the sun-earth distance 
for $\alpha \stackrel{>}{\sim} 10^{-13}$ 
eV$^2$, the corresponding limit on \dm is 
$\Delta m^2 \stackrel{>}{\sim} 10^{-7} $ eV$^2$. 
In an earlier paper we had considered high values of $\Delta m^2$ ($> 10^{-3}
eV^2$) so that the matter effects inside the sun can be neglected and 
one can have decay as well as $\Delta m^2$ independent average oscillations
\cite{cgm}. 
In this paper we consider $10^{-6}\leq \Delta m^2 \leq 10^{-3}$ eV$^2$ 
such that the matter effects inside the sun are important. 
We incorporate the earth matter effects as well.

In section 2 we discuss the usual two flavor MSW solutions to the 
solar neutrino problem. We perform a $\chi^2$-analysis to the 
global data on rates and SK day-night spectrum and present the 
allowed regions in the $\dm-\tan^2\theta$ plane. In section 3 
we introduce the possibility of having  
one of the neutrino states, $\nu_2$ to be unstable. We look 
for the effects of decay on the MSW solutions and present the allowed
areas at various values of the decay constant. We show that for the majoron 
decay model (model 2) the SK data  on $\bar\nue -p$ events
restricts the allowed values of the 
decay constant to extremely low values. 
In section 4 we discuss the implications of non-zero 
neutrino decay for the SNO experiment. We finally present our conclusions 
in section 5.  

\section{MSW effect for stable neutrinos}

As was first indicated in \cite{msw},  interaction of 
electron neutrinos propagating
through the sun modifies the vacuum mixing angle $\theta$ to matter mixing 
angle $\theta_M$ where,
\begin{equation}
\tan 2\theta_{M} =  \frac{\Delta m^2\sin 2\theta}{\Delta m^2\cos2\theta - 
2\sqrt{2}G_{F}n_{e}E}.
\label{thetam}
\end{equation}
Here $n_{e}$ is the ambient electron density, $E$ the
neutrino energy, and \dm (= ${m_{2}^2 - m_{1}^2}$) the mass squared
difference in vacuum.
The vanishing of the denominator in eq. (\ref{thetam}) defines the 
resonance condition, where the matter mixing angle is maximal. 

\noindent
The probability amplitude of survival for an electron neutrino
is given by, 
\begin{equation}
A_{ee} = A_{e1}^\odot  A_{11}^{vac} A_{1e}^\oplus + A_{e2}^\odot A_{22}^{vac}
A_{2e}^\oplus
\label{amp}
\end{equation}
where $A_{ek}^\odot$ gives the probability amplitude of 
$\nu_e \rightarrow \nu_k$ 
transition at the solar surface, 
\begin{equation} 
A_{ek}^\odot = a_{ek}^\odot e^{-i \phi^\odot_{k}}
\eeq
\beq
{a_{e1}^\odot}^2 = P_{J} \sin^2 \theta_M + 
(1 - P_{J}) \cos^2 \theta_M = 1 - {a_{e2}^\odot}^2
\eeq
$P_{J}$ is the non-adiabatic level jumping probability 
between the two mass eigenstates for which we use the standard expression 
from \cite{petcov}.
\beqa
P_J = \frac{\exp(-\gamma \sin^2 \theta) - \exp(-\gamma)}{1-\exp(-\gamma)}
\eeqa
\beqa
\gamma =\pi\frac{\Delta m^2}{E}\left| \frac{d~ln n_e}{dr}
\right |_{r=r_{res}}^{-1}
\eeqa  
$A_{kk}^{vac}$ is the transition amplitude from the solar surface to the 
earth surface,
\beq
A_{kk}^{vac} = e^{-i E_{k} (L - R_{\odot})}
\eeq
where $L$ is the sun-earth distance and $R_\odot$ the radius of the sun. 
$A_{ke}^\oplus$ denotes the $\nu_k\rightarrow \nue$ 
transition amplitudes inside the earth.
We evaluate these amplitudes by assuming the earth to consist of 
two constant density slabs.  
The $\nu_e$ survival probability is given by 
\begin{eqnarray}
P_{ee}  & = &  |A_{ee}|^2 \nonumber \\
        & = & {a_{e1}^\odot}^2 |A_{1e}^\oplus|^2 
+ {a_{e2}^\odot}^2 |A_{2e}^\oplus|^2 
                \nonumber \\
        &   & + 2 a_{e1}^\odot a_{e2}^\odot  
               Re[A_{1e}^\oplus {A_{2e}^\oplus}^{*}
e^{i(E_{2} - E_{1})(L - R_{\odot})} e^{i(\phi_{2,\odot} - \phi_{1,\odot})}]  
\label{pr}
\end{eqnarray} 
It can be shown that the square bracketed term
containing the phases average out to 
zero in the range of \dm in which we are interested  \cite{dighe}.  

Using eq. (\ref{pr}) as the probability we have done a  
$\chi^2$-analysis of the current solar neutrino data on total 
rates from Cl, Ga\footnote{We use the weighted average of the 
rates from Sage, Gallex and GNO.} \cite{solar} and SK
and the 18+18 bins of data on the day-night electron energy 
spectrum from SK\cite{suzuki}.
The details of the code used can be found in \cite{gmr2}. 
We use the 1117 days data of SK and 
incorporate the BBP00 solar model \cite{bp00}.
The theory errors and their correlations in the analysis of 
total rates is included as in \cite{flap}. In addition to the 
astrophysical uncertainties included in \cite{gmr2} 
for the analysis of the total rates, we have included the 
uncertainty in the $S_0$-factor 
for the reaction $^{16}O(p,\gamma)^{17}F$ \cite{bp00}.
For the day-night spectrum analysis we have 
included the correlation between the systematic errors of the 
day and the night bins.
As in \cite{valle} we vary the normalisation of the spectrum as a 
free parameter which avoids the overcounting of the rates and 
spectrum data for SK. Hence for the day-night spectrum analysis 
we have (36 - 1) degrees of freedom (d.o.f) while for the total rates 
we have 3, which makes a total of 38 d.o.f for the rates+spectrum 
analysis with no oscillation. 
For MSW analysis with two active neutrino flavors 
we present in Table 1 two sets of results 
\begin{itemize}
\item with the $^{8}{B}$ normalisation factor $X_B$ 
in the total rates held at the SSM value. 
\item with the $^{8}{B}$ normalisation factor $X_B$ 
in the total rates kept as a free 
parameter. 
\end{itemize}

\begin{tabular}{|c|c|c|c|c|c|}
\hline
&Nature of & $\Delta m^2$ &
$\tan^2\theta$&$\chi^2_{min}$& Goodness\\
&Solution & in eV$^2$&  & & of fit\\
\hline
&SMA & $5.48 \times 10^{-6}$&$5.79 \times 10^{-4}$ &
43.22 & 19.01\%  \\ \cline{2-6}
$X_B$ fixed &LMA & $4.17 \times 10^{-5}$ & 0.35 &
37.33 &  40.78\% \\  \cline{2-6}
at SSM value&LOW & $ 1.51 \times 10^{-7}$ & 0.64 & 39.54 & 31.48\% 
 \\ \hline
&SMA & $5.35 \times 10^{-6}$&$4.34 \times 10^{-4}$ &
37.98 & 33.51\%  \\ \cline{2-6}
$X_{B}$ varying &LMA & $4.22 \times 10^{-5}$ & 0.25 &
34.20 & 50.67\%  \\  \cline{2-6}
&LOW &$1.51 \times 10^{-7}$&$0.64$ &
39.88& 26.20\%  \\ \hline
\end{tabular}

\begin{description}
\item{Table 1:} The best-fit values of the parameters,
$\chi^2_{min}$, and the goodness of fit from the 
global analysis of 
rates and the day-night spectrum data for MSW analysis involving 
two neutrino flavors. 
\end{description}
Thus the best-fit comes in the LMA region.
In fig. 1 we plot the 90\%, 95\%  and 99\% C.L. allowed regions
for the two-flavor MSW transition to an active neutrino.
All the contours have been drawn with respect to the global minimum 
with the $X_B$ fixed at the SSM value.
We have also done a $\chi^2$-analysis for 
$\nu_e - \nu_s$ MSW conversion, $\nu_s$ being a sterile neutrino, 
for both $X_B$ fixed at SSM and $X_B$ varying freely. The 
best-fit values of parameters, $\chi^2_{min}$ and the goodness of 
fit (g.o.f) are 
\begin{itemize}
\item
$\dm = 3.74 \times 10^{-6} $ eV$^2$,~~~$\tan^2 \theta = 5.2 \times 10^{-4}$,
~~~$\chi^2_{min} = 44.85$,~~~g.o.f = 14.79\%,~~~$X_{B}$ fixed at SSM value.

\item 
$\dm = 3.71 \times 10^{-6} $ eV$^2$,~~~$\tan^2 \theta = 4.72 \times 10^{-4}$,
~~~$\chi^2_{min} = 43.42$,~~~g.o.f = 15.53\%,~~~$X_{B}$ free.
\end{itemize}

\section{MSW effect and unstable neutrinos}

If a neutrino of energy $E$ decays while traversing a distance $L$ then 
the decay term $\exp(-\alpha L/E)$ gives the fraction of neutrinos 
that survive, where $\alpha$ is the decay constant discussed before.
In fig. 2 we plot this $\exp(-\alpha L/E)$ as a function of $\alpha$ 
for two different energy values.
For $\alpha$ very small the exponential term is $\approx$ 1 and there is no 
decay. As $\alpha$ increases  
one starts getting decay over the sun-earth distance only for 
$\alpha \stackrel{>}{\sim} 10^{-13}$ eV$^2$.  
As we have discussed in the introduction, since from K-decay 
$\Delta m^2 \geq 10^{6}\alpha$, one can have simultaneous MSW 
and decay for $10^{-7}$ eV$^2 < \dm < 10^{-3}$ eV$^2$. 
Below this \dm
the bound on the coupling constant coming from K-decay restricts the 
decay constant $\alpha$ to be small enough so that decay effects are 
negligible over the sun-earth distance. 
So henceforth we will be concerned with only the LMA and SMA 
solutions, the LOW region remaining unaffected due to decay. 
For $\Delta m^2 > 10^{-3}$ eV$^2$ there will not be any MSW effect. 

The probability amplitude for $\nue$ survival 
in presence of neutrino decay is again given by
eq.(\ref{amp}), with $A_{ek}^\odot$, $A_{ke}^\oplus$ and $A_{11}^{vac}$ 
as before, the only change being for $A_{22}^{vac}$ since the 
$\nu_2$ decays on its way from the sun to earth. $A_{22}^{vac}$ 
is given by 
\beq
A_{22}^{vac} =  
e^{-i E_{2} (L - R_{\odot})} e^{-\alpha (L - R_{\odot})/E_2}
\eeq
Then, the $\nu_e$ survival 
probability is given by (ignoring the phase part), 
 
\begin{eqnarray}
P_{ee}  & = & {a_{e1}^\odot}^2 |A_{1e}^\oplus|^2 + 
{a_{e2}^\odot}^2 |A_{2e}^\oplus|^2 e^{-2 \alpha
              (L - R_{\odot})/E_2}  
\label{prdecay}
\end{eqnarray} 
The day-time probability (i.e. without the earth effect) is given by
\begin{eqnarray}
P_{ee}^{\rm day} & = & \cos^2 \theta [ P_J \sin^2 \theta_M + (1 - P_J) \cos^2 \theta_M]
                \nonumber \\
       &  & + \sin^2 \theta [ (1 -P_J) \sin^2 \theta_M + \cos^2 \theta_M P_J]
              e^{-2 \alpha (L - R_{\odot})/E_2}
\label{peesimple}
\end{eqnarray}
From eq.(\ref{peesimple}) we note that the
the decay term appears with a $\sin^2\theta$ and is therefore 
appreciable only for large enough $\theta$. Thus we expect
the effect of decay to be maximum in the LMA region. 
This can also be understood as follows. 
The $\nu_e$ are produced mostly as $\nu_2$ in the solar core. 
In the LMA region the neutrinos move adiabatically through the sun and 
emerge as $\nu_2$ which eventually decays. For the 
SMA region on the other hand $P_J$ is non-zero and $\nu_e$
produced as $\nu_2$ cross over to $\nu_1$ at the resonance and come out 
as a $\nu_1$ from the solar surface. 
Since  $\nu_1$ is stable, decay does not affect this region. 
Including the earth effect the survival probability can be expressed as
\begin{equation}
P_{ee} = P_{ee}^{\rm day} + \frac{(\sin^2 \theta - P_{2e})(P_{ee}^{\rm day}
+e^{-2 \alpha (L - R_{\odot})/E_2} (P_{ee}^{\rm day} -1))}{cos^2 \theta - 
sin^2 \theta e^{- 2 \alpha (L - R_{\odot})/E_2}}
\label{dn}
\end{equation}
$P_{2e}$ is the probability of the second mass eigenstate to  get 
converted into 
the $\nu_e$ state at the detector. 

In fig. 3 the solid lines show the 
survival probability versus neutrino energy for 
three different values of $\alpha$ with \dm and $\tan^2\theta$ fixed at the 
best fit value of the LMA solution. 
The introduction of decay reduces the survival probability, as the 
second term in eq.(\ref{prdecay}) reduces with $\alpha$. Decay also 
brings about a deviation from a flat probability
towards the high energy end.
The dotted line gives 
the probability for the same \dm but a higher value of $\tan^2\theta$ 
with $\alpha=10^{-11}$ eV$^2$. We observe that there is a huge drop in the 
survival probability initially, followed by a sharp increase with energy.  
A comparison between the different curves shows that inclusion of 
decay results 
in the reduction of the survival probability as well as an energy distortion 
and both these effects increase with the mixing angle $\theta$. 

We next perform a $\chi^2$-analysis of the total rates and the day-night 
spectrum data for non-zero decay.
The best-fit comes for $\alpha = 0$, corresponding to no decay with 
the \dm and $\tan^2\theta$ as given in Table 1. 
However non-zero values of $\alpha$ giving finite decay probability 
also gives acceptable fit.
For small values of $\theta$ as in the SMA region, 
the fraction of $\nu_2$ in the solar neutrino beam is very small. As a 
result very few neutrinos decay and all   
values of $\alpha$ may be allowed.  
On the other hand in the LMA region the $\theta$ is high and 
large number of neutrinos can decay producing a distortion in the 
energy spectrum. Therefore the data can put some restrictions 
on the allowed values of $\alpha$.   
In fig. 4 we plot $\Delta \chi^2 (=\chi^2 - \chi^2_{min})$ vs $\alpha$ 
keeping the other two parameters free (in the LMA regime). 
We also show the 
99\% C.L. limit for three parameters by the dotted line. 
From fig. 4 we see that in the LMA region, values of $\alpha$ upto 
$3.5 \times 10^{-11}$ eV$^2$ 
are allowed at 99\% C.L. from the global analysis. 

In fig. 5 we show the allowed regions in the $\dm -\tan^2\theta$ plane 
for various fixed values of $\alpha$. 
The first panel is 
for $\alpha=0$ which is the case of no decay 
and the contours are the same as those presented in fig. 1, modulo 
the difference in the definition of the C.L. as now there are three  
parameters whereas in fig. 1 we had two. The other three panels 
are for different non-zero allowed values of $\alpha$ obtained from 
fig. 4. 

As expected, we find that the SMA allowed regions 
remain mostly unaffected due to a non-zero $\alpha$. 
The LMA solution on the other hand is appreciably 
affected and the allowed area shrinks as $\alpha$ increases. 
This effect is seen to be more pronounced 
for larger $\tan^2\theta$. The reason 
for this can be traced down to the fact that decay results 
in distortion of the high energy end of the neutrino 
spectrum and this 
distortion is more for higher $\tan^2\theta$, a feature explicit in  
fig. 3. As a result higher values of $\tan^2\theta$ are disfavored 
by the global data as $\alpha$ increases.  

It is to be noted that although in the LOW region the \st is high, the 
\dm is small and appreciable decay over the sun-earth distance is 
not obtained if one has to be consistent with the bounds on the 
coupling constant from K-decays. 
Because of this the LOW region is not plotted in fig. 5. 


The results presented in this section are 
in general valid for both 
the  decay models
though in model 2 
the $\nu_2$ decays to a $\bar{\nu}_1$, which can interact with the electrons
in SK detector as $\bar{\nu}_e$ and $\bar{\nu}_x$.
But since  
the $\bar{\nu}_1$ is degraded in energy \cite{app}. 
the effect of these additional 
($\anue-e$) and ($\bar\nu_x-e$) 
scattering events in SK is not significant \cite{cgm} and 
the final results remain the same. 
However additional constraint on model 2 come from the 
$\bar{\nu_e} p \rightarrow e^{+} n$ events, which contribute 
to the background in SK. From the 
absence of a significant contribution above the background,
the bound obtained on the total flux of $\bar{\nu_e}$ from $^{8}{B}$
neutrinos is $\Phi_{\bar{\nu_e}}$ $(^{8}{B}) < 
1.8 \times 10^5$ cm$^{-2}$ s$^{-1}$ 
which translates to a bound on the probability $P_{e\bar{e}} < 3.5 \%$ 
\cite{lujan}. 
In fig. 6 we plot the conversion probability 
$P_{e\bar{e}}$ vs $\alpha$. 
For each $\alpha$ we find the $\chi^2_{min}$ and 
plot $P_{e\bar{e}}$ for the corresponding 
the best fit \dm and $\tan^2\theta$. 
We also show the allowed 
limit of $P_{e\bar{e}}$ = 0.035. So from this constraint for model 2 
only $\alpha \leq 5.8\times 10^{-13}$ eV$^2$ remain allowed.

\section{Implications for SNO}

The main detection processes in the heavy water of SNO are
\begin{eqnarray}
\nu_e + d \rightarrow p + p +e^-
\label{nued}
\end{eqnarray}
\begin{eqnarray}
\nu_x + d \rightarrow p + n + \nu_x
\label{nuxd}
\end{eqnarray}
The first is a charged current (CC) reaction with 
energy threshold of 1.44 MeV  
and the second  
one is a neutral current (NC) process with an energy 
threshold of 2.23 MeV. While the CC is sensitive to only $\nue$, 
the NC process is sensitive to neutrinos and antineutrinos of all
flavors. 
The rate of ($\nue-d$) CC events recorded in 
the detector is given by
\beqa
R_{CC} = \frac{\int dE_\nu \lambda_\nue(E_\nu)
\sigma_{CC}(E_\nu)\langle P_{ee}\rangle}
{\int dE_\nu \lambda_\nue(E_\nu)
\sigma_{CC}(E_\nu)}
\eeqa
\beqa
\sigma_{CC} = \int_{E_{A_{th}}}dE_A\int_0^\infty
dE_TR(E_A,E_T)\int dE_\nu \frac{d\sigma_{\nue d}
(E_T,E_\nu)}{dE_T}
\label{ccosc}
\eeqa
where 
$\lambda_\nue$ is the normalised $^8B$ neutrino spectrum, 
$\langle P_{ee}\rangle$ 
is the time averaged $\nu_e$ survival probability, 
d$\sigma_{\nue d}$/dE$_T$ is the differential cross section of the 
($\nu_e-d$) interaction \cite{nued}, 
$E_T$ is the true and $E_A$ the apparent(measured) kinetic energy of
the recoil electrons, $E_{A_{th}}$ is the detector 
threshold energy which we take as 5 MeV  
and R($E_A$,$E_T$) is the energy resolution function which is 
assumed to be Gaussian
\beqa
R(E_A,E_T)=\frac{1}{\sqrt{2\pi}(0.348\sqrt{\frac{E_T}{{\rm MeV}}})}
\exp\left(-\frac{(E_T-E_A)^2}{0.242E_T MeV}\right)
\label{ncosc}
\eeqa
The NC event rate is given by
\beqa
R_{NC}= \frac{\int dE_\nu\lambda_\nue(E_\nu)\sigma_{N
C}(E_\nu) {\cal P}}
{\int dE_\nu\lambda_\nue(E_\nu)\sigma_{NC}(E_\nu)}
\eeqa
%
where  
${\cal P} 
= \langle P_{ee}\rangle +\langle P_{e\mu}\rangle$ for two flavor 
$\nue-\numu$ oscillation. 
The left hand panels in 
fig. 7 give the range of predicted values of 
$R_{CC}$, $R_{NC}$ and the double ratio $R_{NC}$/$R_{CC}$ at 99\% C.L., 
from the global MSW analysis of the solar neutrino data. 
The black dots in these figures represent the 
expected rates in SNO for the local best 
fit values of the parameters obtained. 
As the neutral current interaction is flavor blind, 
the $R_{NC}$ remains 1 for oscillations to active neutrinos for which  
${\cal P}=1$. But for oscillations to sterile neutrinos 
${\cal P}=\langle P_{ee}\rangle$ and $R_{NC}=0.78$ for the best fit 
case. 

The right hand panels in fig. 7 show the 
the corresponding rates in SNO for the 
LMA region if the neutrinos are assumed to be unstable (the SMA solution 
remains largely unaffected as discussed in the previous section). 
We consider the decay model 1 and present the  
99\% C.L. predicted range of values of the rates  
for three different allowed values of the parameter  
$\alpha$, along with the case for $\alpha=0$, which is the same as the 
LMA region of global MSW analysis, modulo the difference in the 
definition of the C.L.. Also shown are the rates for the best fit value 
of \dm and $\tan^2\theta$ for various fixed values of $\alpha$. 
For each fixed value of \dm and $\tan^2\theta$ one expects 
$R_{CC}$ to fall sharply with alpha. But since the best fit of   
the global analysis for increasing $\alpha$ corresponds 
in general to a higher \dm, that raises the value of $R_{CC}$. 
Hence $R_{CC}$ in SNO alone may not be able to distinguish  
the decay scenario from MSW conversion to active stable neutrinos.
The value of $R_{NC}$ suffers a           
marked decrease from 1
as the decay constant $\alpha$ increases, 
since for the decay model 1 
the $\nue$ decays to sterile particles 
which cannot be detected in SNO. 

From fig.7 we see that if $R_{NC}$ is below 1,
one can 
have either MSW conversion to sterile neutrinos, or the possibility of 
unstable neutrinos. But the ambiguity between the last two 
cases can be lifted if one next looks at the value of $R_{CC}$, 
which is much lower for the case of unstable neutrinos than what one 
would get for MSW conversion to sterile neutrinos. 
We also display the expected range of $R_{NC}$/$R_{CC}$ 
for various fixed values of the decay constant. It is clear from 
the figure that from the value of $R_{NC}$/$R_{CC}$ in SNO one can clearly 
distinguish the case of decay from the case of MSW transition to sterile 
neutrinos.
  
In the decay model 2 the $\nue$ decay to $\bar\nu_1$, which 
can in principle show up both in the NC events in SNO as well as 
in the charged current absorption reaction  
\begin{eqnarray}
\anue + d \rightarrow n + n +e^+
\label{anued}
\end{eqnarray}
But since the absence of ($\anue-p$) events in SK restrict 
$\alpha \leq 5.8\times 10^{-13}$ eV$^2$, both the additional contribution 
to the NC event as well as the ($\anue-d$) absorption is negligible 
in SNO. 

      
\section{Conclusions}

We have investigated the effect of neutrino decay on the MSW solution 
to the solar neutrino problem. 
There are two factors which control the effect of decay
\begin{itemize}
\item The mixing angle $\theta$ which determines the fraction of the 
unstable component in the $\nue$ beam
\item The decay constant $\alpha$ which determines the decay rate
\end{itemize}
In order for the neutrinos to decay over the earth sun distance 
$\alpha$ should be $\geq 10^{-13}$ eV$^2$. Then eq.(\ref{galpha})
and the bounds on the coupling constant from K decays  
restricts $\Delta m^2 \stackrel{>}{\sim} 10^{-7}$ eV$^2$ and 
decay does not take place in the LOW region.
We therefore probe the SMA and the LMA regions.   
We find that even the SMA region is not affected 
much because the mixing angle being small very few neutrinos 
would decay.
But the effect of decay on the LMA region is significant. 
This is borne out by the 
$\chi^2$-analysis 
of the global solar data on rates and SK day-night spectrum. We 
point out that although the data {\it prefers} no decay, it still 
{\it cannot rule out} the possibility of unstable neutrinos completely. 
We put limits on the allowed range of the decay constant $\alpha$ and 
present the allowed areas in the $\Delta m^2-\tan^2\theta$ 
parameter space for various allowed values of $\alpha$. The LMA 
allowed zone is seen to be severely constrained by decay. 

From fig. 4, 
values of $\alpha \leq 3.5\times 10^{-11}$ eV$^2$ are allowed, 
implying a neutrino 
rest frame lifetime $\tau_0 \geq 2 \times 10^{-5}$ sec. Thus 
one may encounter decay before neutrino decoupling in the 
early universe. This may result in increasing the number of light 
neutrinos $N_\nu$ to greater than 3. However, the upper limit on 
$N_\nu$ can be as large as 6 \cite{cosmo}. Hence our decay model 
is consistent with the bounds from early universe.  

From the fact that the neutrinos from SN1987A have not decayed on their way 
one gets a lower bound 
on the electron neutrino lifetime as $\tau > 5.7 \times 10^{5} (m_{\nu_e}$/eV)
sec. However if one includes neutrino mixing then shorter lifetimes are 
allowed provided $|U_{e2}| < 0.9$ \cite{haber}. This again is consistent 
with our analysis.

In this paper we assume one of the neutrino states to be unstable 
in vacuum and the values of the decay constant  
$\alpha$ are such that one has decay over earth-sun distance. 
For these values of $\alpha$ 
decay inside the sun or the earth is negligible. However as was 
shown in \cite{bzv} matter can induce neutrino decay with majoron 
emission even when neutrinos are stable in vacuum. 
It was shown later in \cite{bzrossi} that in the presence of only 
standard interactions,
the matter induced decay cannot provide a solution to the
solar neutrino problem  
due to strong bounds on neutrino-majoron 
couplings. 
As we have discussed  in this paper,  decay models where the $\nu_2$ decays
to a $\bar\nu_1$ and a majoron is independently 
disfavored from the non-observation
of ($\anue-p$) events in SK.

We have looked at the implications of unstable neutrinos 
for the SNO detector.
We first present the values of the CC and NC rates 
which SNO is expected to observe for MSW transitions with stable 
neutrinos. We have given the values for transitions to both 
active as well as sterile species. 
For unstable neutrinos the NC rate is less than 1 and  
comparable 
to $R_{NC}$ for standard MSW SMA transition to sterile neutrinos. 
But even 
though the value of $R_{NC}$ may be nearly same for the two scenarios, 
comparing the values of $R_{CC}$ or $R_{NC}/R_{CC}$ it may be possible 
in principle to distinguish the decay scenario 
from the MSW  transition to  sterile neutrinos.

{\small The authors wish to thank Carlos Pe\~{n}a-Garay for many 
useful discussions during the development of the solar code, 
Y. Suzuki for sending them the SK data and H.M. Antia for many 
helpful discussions.}



\centerline{\psfig{figure=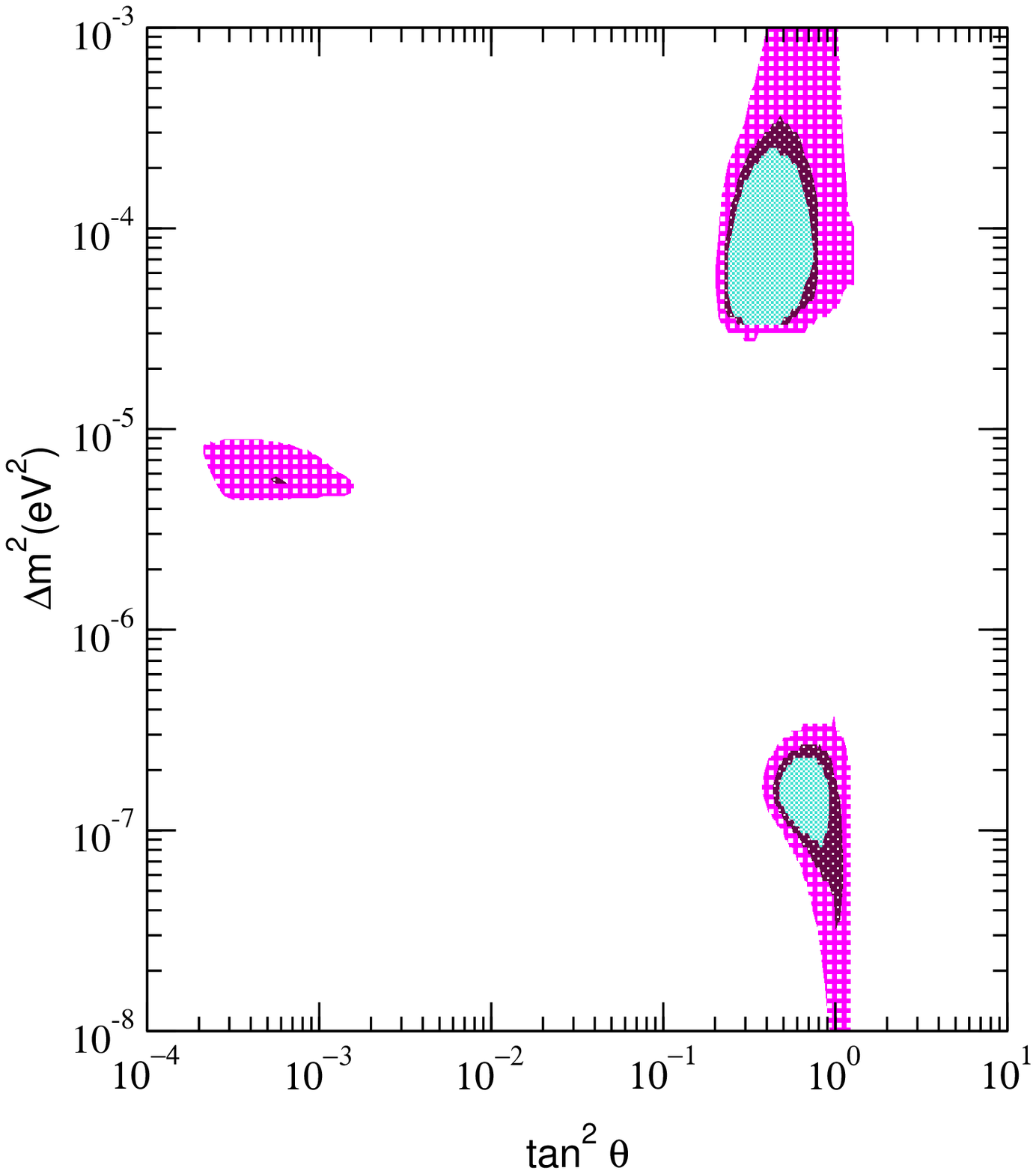,width=16.5cm,height=20.5cm}}
\vskip -1in
\parbox{5in}{
{\bf Fig. 1}: The 90, 95 and 99\% C.L. allowed area from the 
global analysis of the total rates from Cl, Ga and SK detectors 
and the 1117 days SK recoil electron spectrum at day and night, 
assuming MSW conversions to stable sequential neutrinos.}

\centerline{\psfig{figure=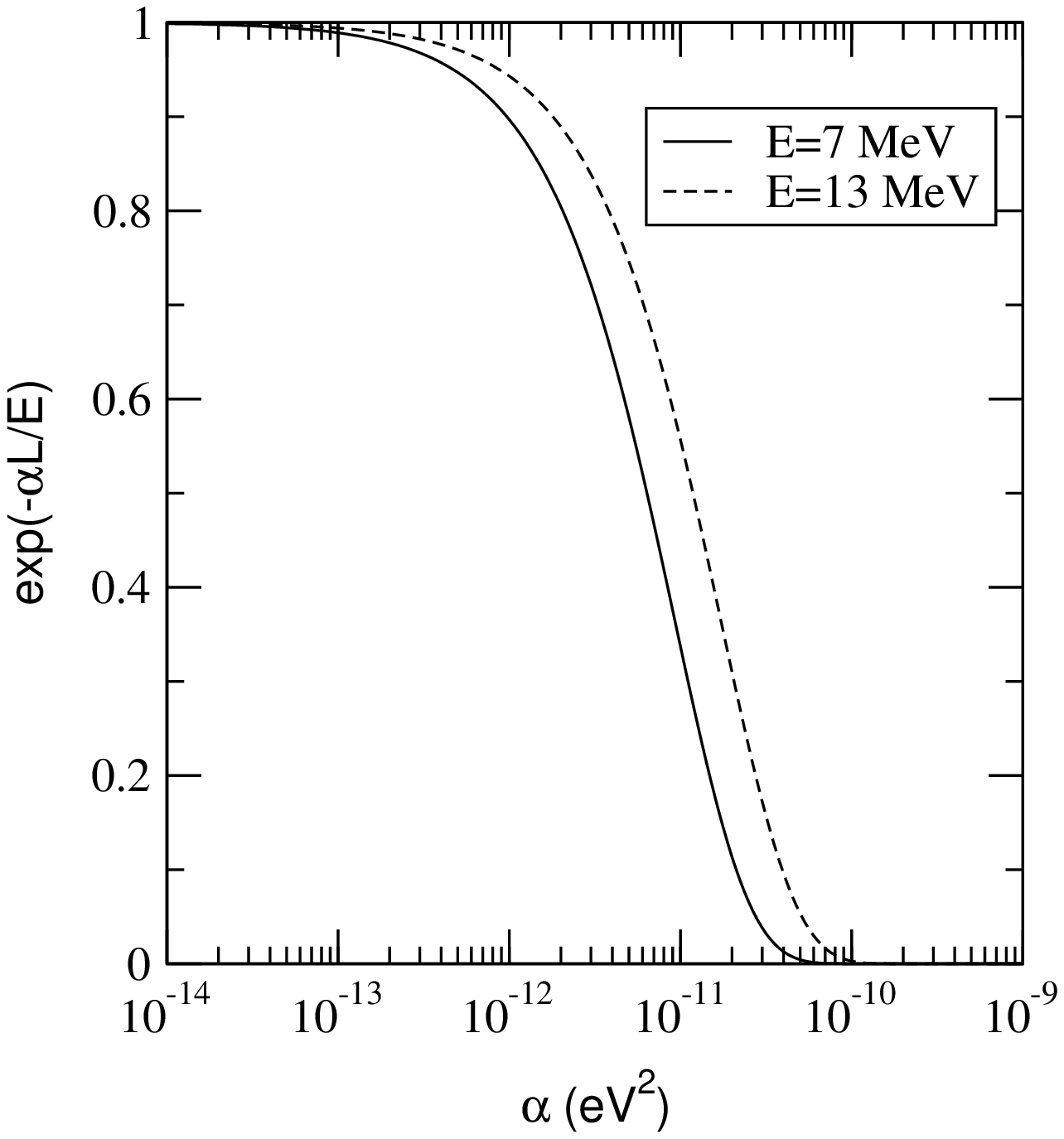,width=17.5cm,height=20.5cm}}
\vskip -2.5in
\parbox{5in}{
{\bf Fig. 2}: The decay term $\exp(\alpha L/E)$ vs $\alpha$ 
for two different neutrino energies. $L$ is taken as the earth-sun 
distance.}

\centerline{\psfig{figure=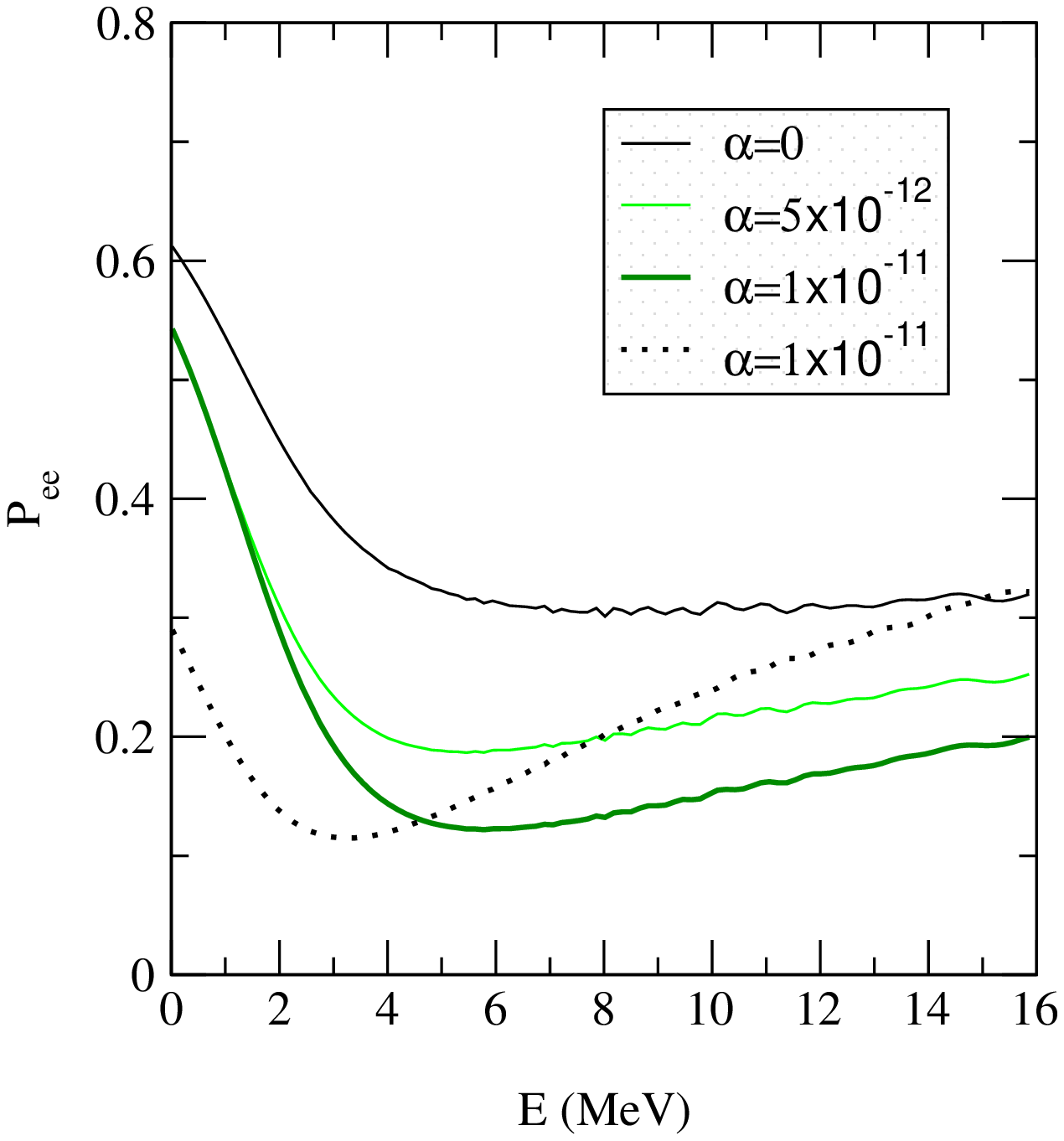,width=17.5cm,height=20.5cm}}
\vskip -2.5in
\parbox{5in}{
{\bf Fig. 3}: The survival probability as a function of 
$\nue$ energy for different values of $\alpha$. The solid lines 
are for $\tan^2\theta=0.35$ while the dotted line is for 
$\tan^2\theta=0.85$. For all the curves 
$\Delta m^2=4.17\times 10^{-5}$ eV$^2$.}

\centerline{\psfig{figure=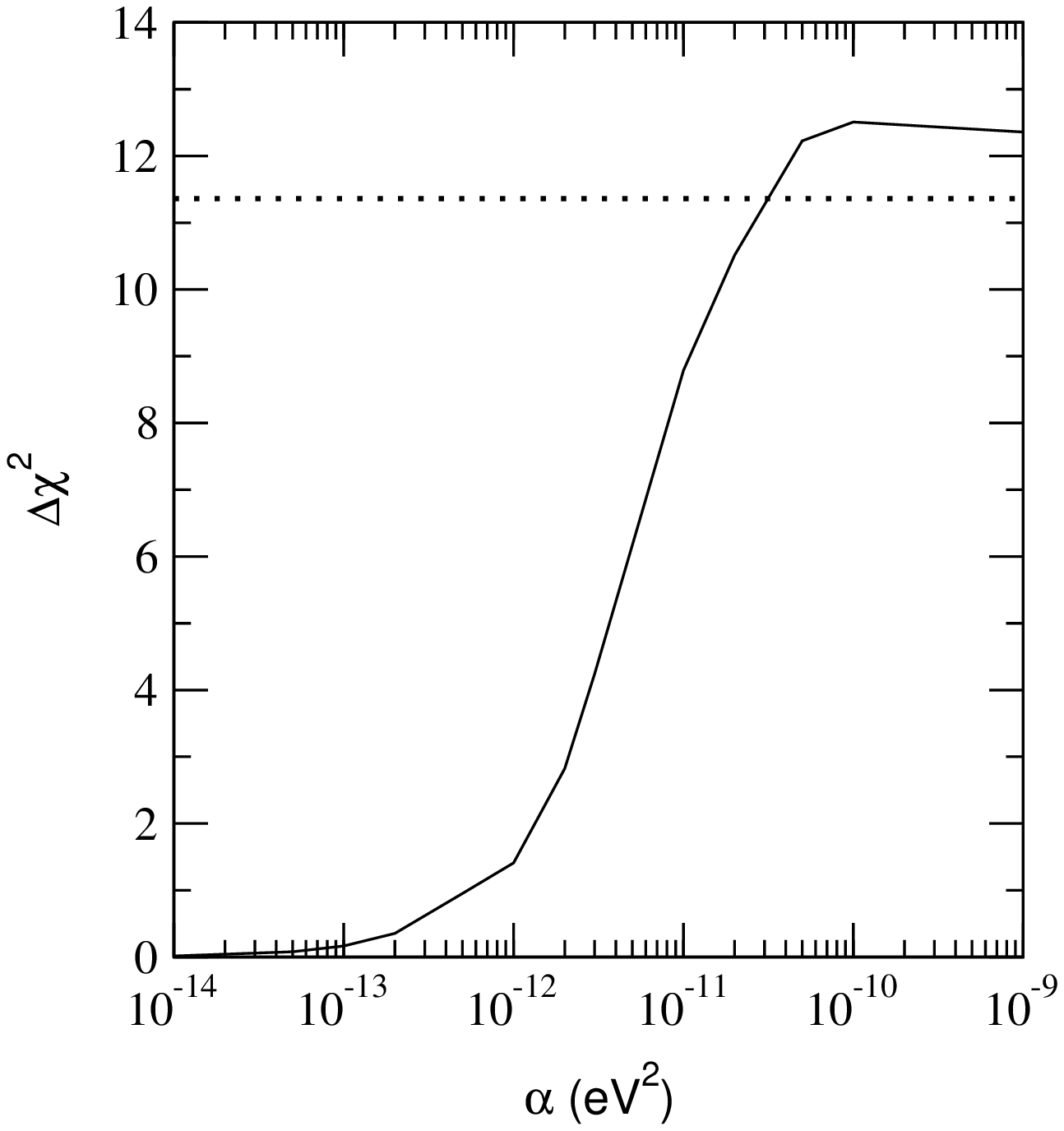,width=17.5cm,height=20.5cm}}
\vskip -2.5in
\parbox{5in}{
{\bf Fig. 4}: The $\Delta \chi^2$ versus decay constant $\alpha$ for the 
global analysis of total rates and the SK day and night spectrum data. 
Also shown by the dotted line is the 99\% C.L. limit for three parameters.} 

\centerline{\psfig{figure=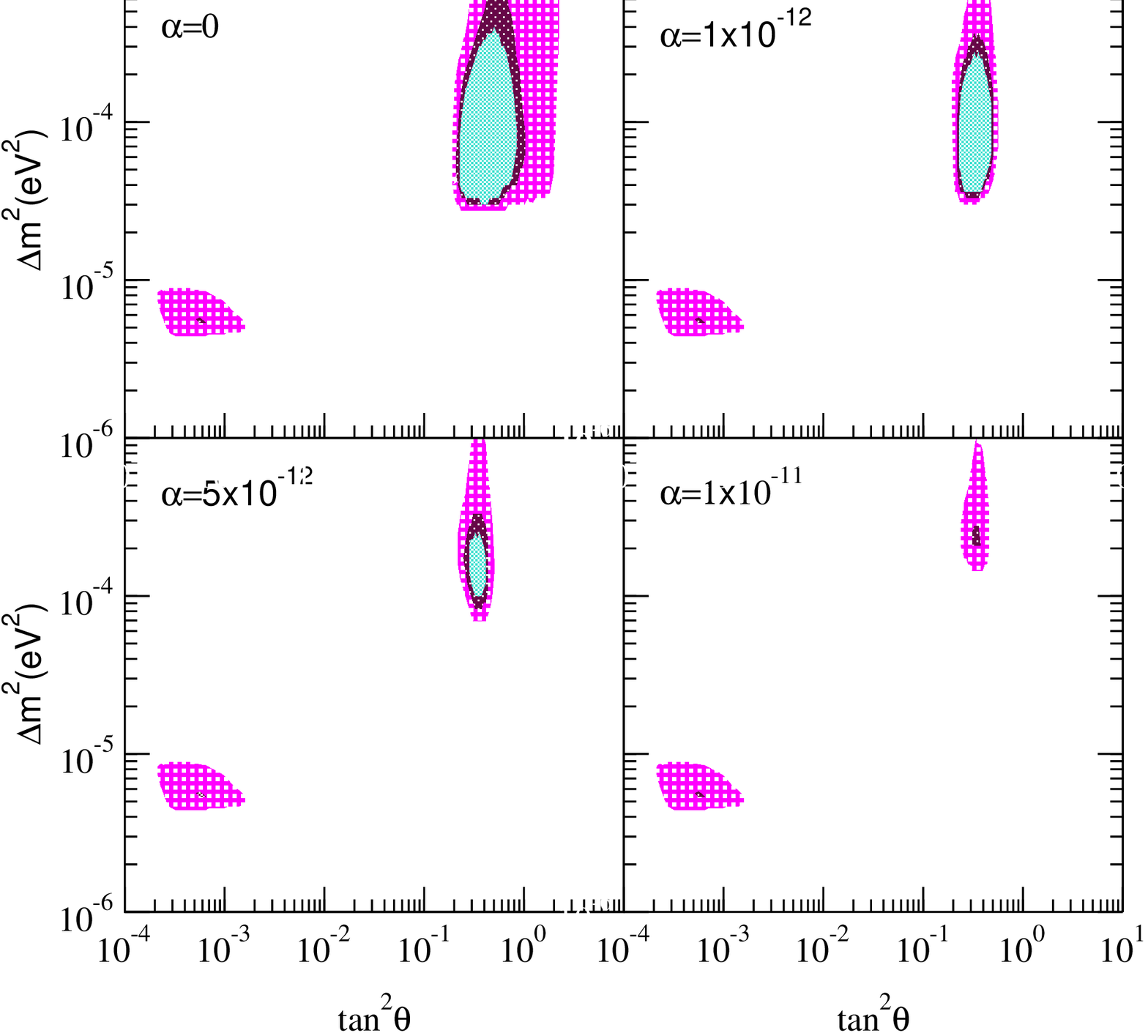,width=16.5cm,height=20.5cm}}
\vskip -2.5in
\parbox{5.5in}{
{\bf Fig. 5}: The 90, 95 and 99\% C.L. allowed area from the
global analysis of the total rates 
and the 1117 days SK day-night spectrum data,
for MSW conversion of unstable neutrinos at various values of the decay 
constant $\alpha$. The different fixed values of $\alpha$ in eV$^2$ are 
indicated. }

\centerline{\psfig{figure=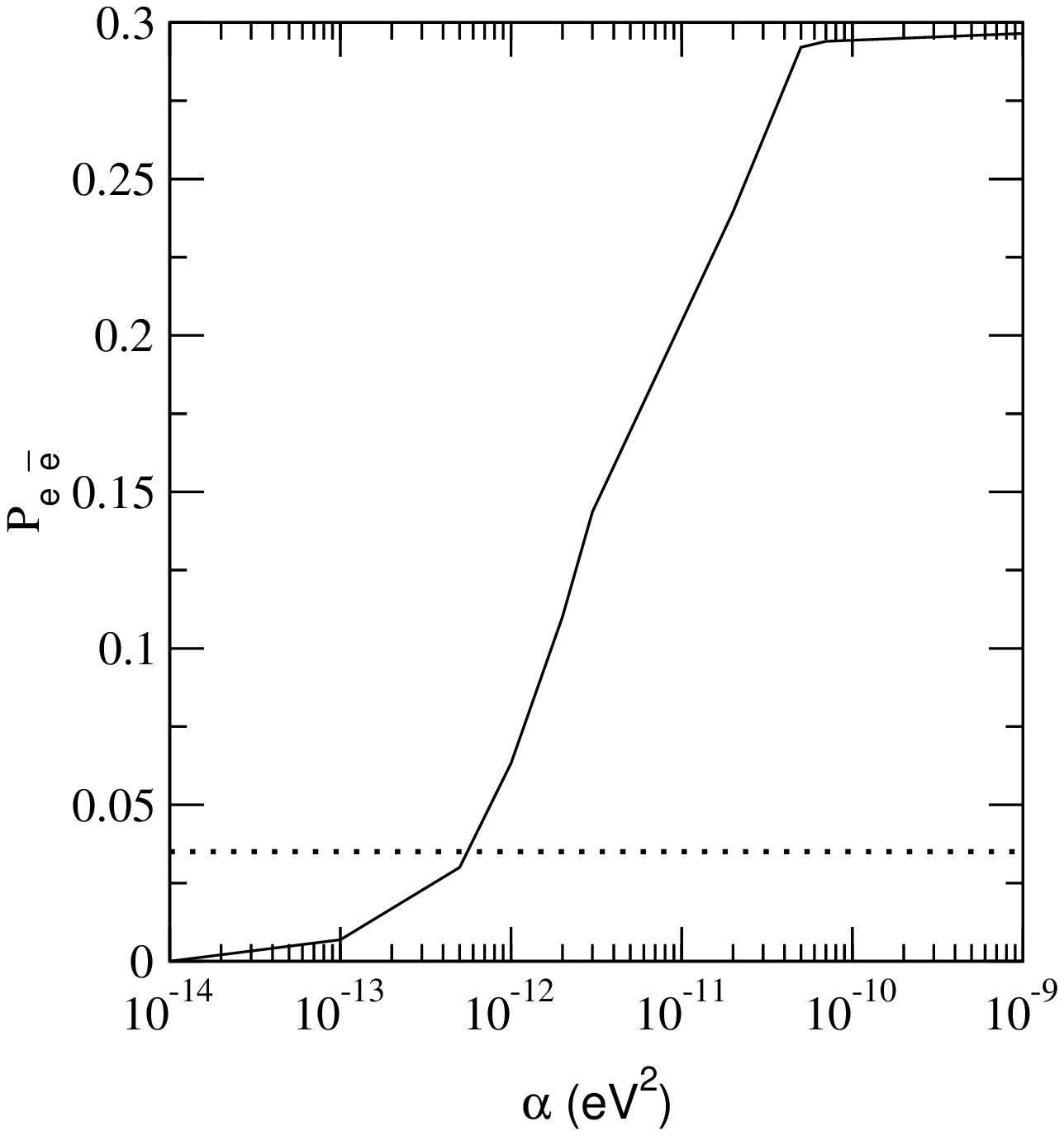,width=16.5cm,height=20.5cm}}
\vskip -2.5in
\parbox{5in}{
{\bf Fig. 6}: The conversion probability of $\nue$ to $\anue$, 
$P_{e\bar{e}}$ versus decay constant $\alpha$. 
Also shown by the dotted line is the 99\% C.L. value of $P_{e\bar{e}}$ 
allowed from the 
non-observance of ($\anue-p$) events in the SK detector.} 

\centerline{\psfig{figure=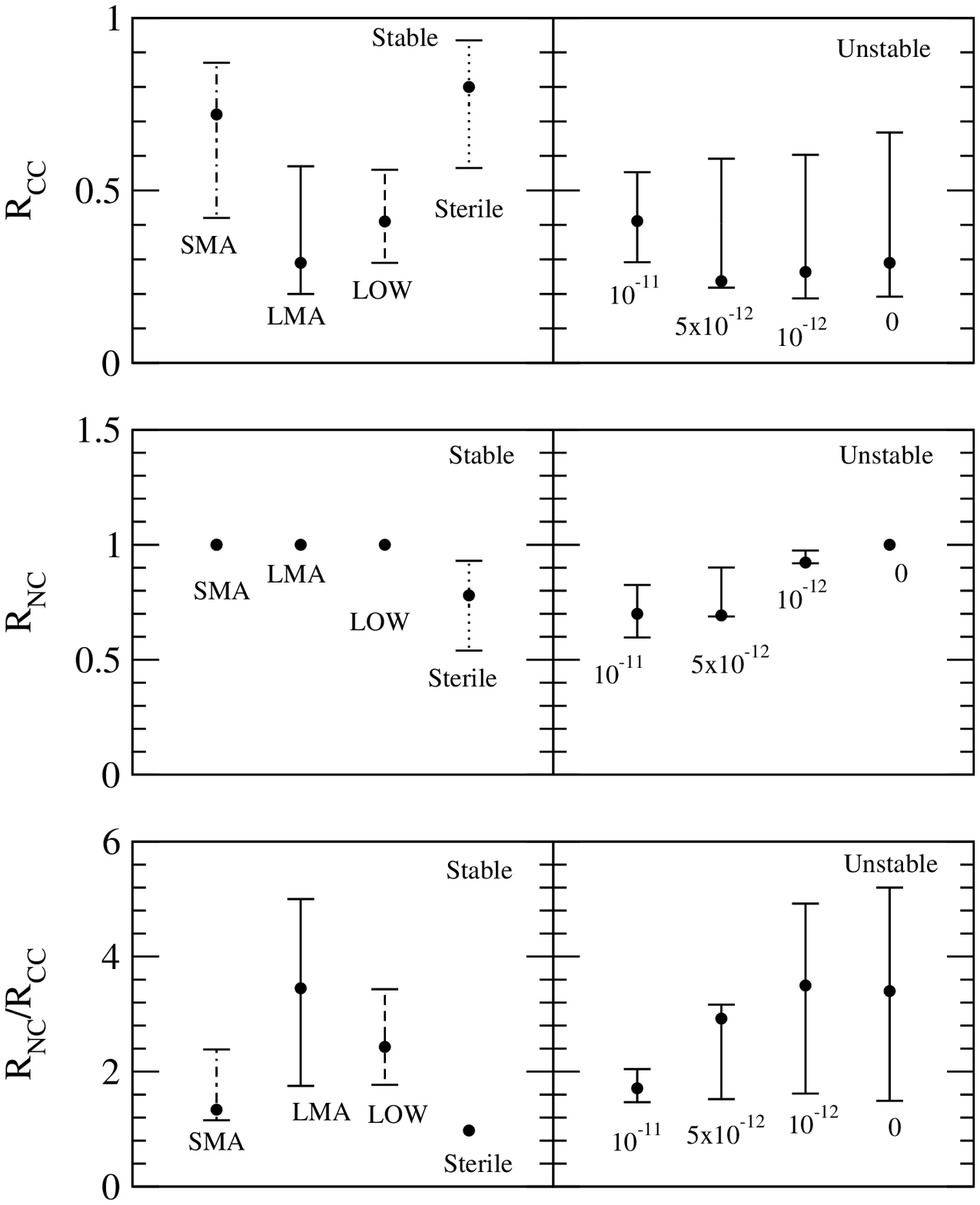,width=16.5cm,height=20.5cm}}
\vskip -1.0in
\parbox{5in}{
{\bf Fig. 7}: The predicted ranges of $R_{CC}$, $R_{NC}$ and 
$R_{NC}/R_{CC}$ in SNO for the standard MSW conversion to stable 
neutrinos (left hand panels) and for unstable neutrinos (right hand 
panels). The dots give the values of the rates for the local best fit.}

\end{document}